\documentclass[prb,twocolumn,showpacs]{revtex4}
\usepackage{amsfonts}
\usepackage{amsmath}
\usepackage{amssymb}
\usepackage{graphicx}
\usepackage{epstopdf}

\begin{document}

\title{Intrinsic high-temperature superconductivity in ternary iron selenides%
}
\author{Shin-Ming Huang$^{1}$, Chung-Yu Mou$^{1,2,3}$, and Ting-Kuo Lee$^{2}$
}
\affiliation{$^{1}$Department of Physics, National Tsing Hua University, Hsinchu 30043,
Taiwan}
\affiliation{$^{2}$Institute of Physics, Academia Sinica, Nankang, Taiwan}
\affiliation{$^{3}$Physics Division, National Center for Theoretical Sciences, P.O. Box
2-131, Hsinchu, Taiwan}

\begin{abstract}
We examine superconductivity in the mesoscopically mixed antiferromagnetic
(AF) and superconducting (SC) phases of ternary iron selenides K$_{y}$Fe$
_{2-x}$Se$_{2}$. It is shown that the interlayer hopping and AF order are
key factors to determine $T_{c}$ of the SC phase. In general, the hopping
will produce deformed Fermi surfaces (FS's) that tend to suppress
superconductivity. However, contrary to the common expectation, we find that
larger AF order actually results in larger SC order, which explains the
observed relatively high $T_{c}$ in these phases. Furthermore our results
indicate that by reducing the interlayer hopping appropriately,
phase-separated K$_{y}$Fe$_{2-x}$Se$_{2}$ may exhibit its intrinsic SC phase
in the two dimensional limit with a much higher $T_{c}$ ($\sim 65K$) than
what has been observed.
\end{abstract}

\maketitle

\section{Introduction}

The newly discovery of ternary iron-selenide superconductors A$_{y}$Fe$%
_{2-x} $Se$_{2}$ (A=K, Rb, Cs, and Tl) \cite{JGuo, AFWang} opens an
interesting route to explore the origin of high temperature
superconductivity in Fe-based superconductors. These materials have $T_{c}$
up to 30K, which is relatively high in comparison to the average $T_{c}$ in
the family of Fe-based superconductors. However, unlike many other Fe-based
materials, in which the SC order gets suppressed in the presence of the AF
order due to their strong competition, early resistivity measurements \cite
{JGuo,Fang2011,DMWang2011} surprisingly found that AF order and SC
order coexisted while $T_{c}$ was still kept relatively
high. Further systematic investigations reveal that $T_{c}$'s and the AF
transition temperatures ($T_{N}$) of these materials exhibit similar trends:
Both $T_{N}$ and $T_{c}$ are higher in the SC samples than those in the
non-SC sample. This clearly implies that the coexisted
antiferromagnetic ordering and superconductivity are not simply competing against each other
\cite{Liu2011}. In addition, it is found that the AF phase coincides with
the $\sqrt{5}\times \sqrt{5}$ Fe-vacancy order with extraordinarily large
magnetic moment of 3.3$\mu _{B}$/Fe \cite{Bao2011}. These results prompt a
critical examination and explanation on how the AF order with large moment
can coexist with the SC order while $T_{c}$ remains so high.

To explore the origin of relatively high $T_{c}$ in ternary iron-selenides,
the nature of the phase with coexistence of SC and AF orders is further
examined. A number of experiments \cite{ZWang2011, FChen2011, Texier2012,
Charnukha2012, Shermadini2012, Stadnika2013} show that instead of being
coexistent homogeneously, the SC and AF orders are phase separated at
mesoscopic scales. In particular, the volume fraction of the SC phase is
estimated to be less than 20\% by using local probes \cite%
{Texier2012,Shermadini2012}. Furthermore, it is shown that metallic behavior
is exhibited in the SC phase \cite{Texier2012}, while semiconducting
behavior is found in the magnetic phase \cite{Charnukha2012}. In addition, a
heterostructure arrangement of SC and AF layers stacking alternatively is
observed in TEM experiments \cite{ZWang2011}, in consistent with the picture
suggested by Charnukha \textit{et al}. \cite{Charnukha2012}. A more direct
visualization is obtained by recent STM results \cite{Li2012}, in which two
distinct regions along c-axis are clearly identified in K$_{x}$Fe$_{2-y}$Se$%
_{2}$ compound, SC KFe$_{2}$Se$_{2}$ (122 system) and insulating K$_{x}$Fe$%
_{1.6}$Se $_{2}$ (245 system) with the $\sqrt{5}\times \sqrt{5}$ order.
Furthermore, it is found that pure KFe$_{2}$Se$_{2}$ could exist in a
metallic state without superconductivity but with weak charge density wave
\cite{Xue2012}.

On the theoretical side, much work has been devoted to understand
homogeneous AF and SC phases of either 122 or 245 system \cite
{Yan2011,Cao2011,WLi2012,Liu2012,
Maiti2011,Maier2011,YZhou2011,Das2011,Huang2011,Huang2012}. Little is known
about the mechanism of superconductivity in the combined system. Recently,
Jiang \textit{et al}. \cite{Jiang2012} investigated a bilayer
heterostructure with both SC and AF phases. Based on the pair-hopping
approximation between the SC and AF layers, it is shown that a drop of
magnetic moment occurs in the AF layer when the temperature goes below the
SC transition, in agreement with the observation of neutron scattering
experiments \cite{Bao2011}. Since the effect on SC phase due to the AF order is found to be
quite substantial \cite{Li2012,Xue2012}, it is interesting to examine what
is the effect of AF order on superconductivity, especially in the presence
of such strong AF order in the iron-vacancy-ordering phase.

In this work, we investigate superconductivity in a bilayer system with iron
vacancy-free layer on top of an iron vacancy-ordered AF layer (245). The
iron vacancy-free layer is nominally taken to be the 122 system with fitted
band structures. The electronic structures are examined under different
strength of interlayer hopping, interlayer spin coupling, and AF order. It is shown that
both the interlayer hopping and interlayer spin coupling generally suppress superconductivity.
In particular, the interlayer hoppinng would result in deformed Fermi surfaces structures
that tend to frustrate the coupling of SC orders on Fermi surfaces. However, 
unexpectedly we find that for fixed hopping amplitude,
larger AF orders actually result in larger SC orders, which explains the
observed relatively high $T_{c}$ and trends of $T_{c}$ versus the AF
transition temperature $T_{N}$ \cite{Liu2011} in these phases. 
Our results imply that in the 2D limit, the pure SC phase in
phase-separated ternary iron selenides may have a much higher $T_{c}$, being
around $65K$.

\section{Theoretical Model}

\begin{figure}[tbp]
\begin{center}
\includegraphics[height=1.92in,width=3.06in] {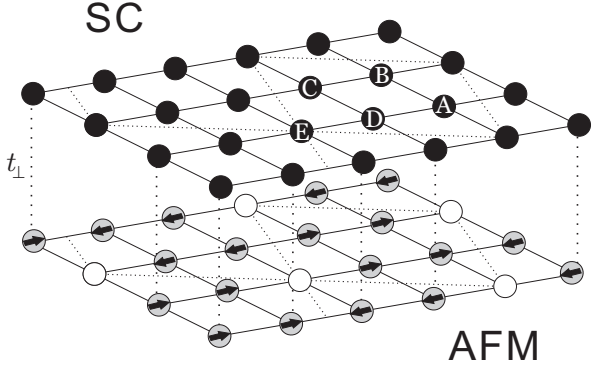}
\end{center}
\caption{Coupled 122-245 bilayer junction. Top lattice is the SC 122 layer
and bottom one is the AF 245 layer. The dotted lines enclose the unit cell
in each layer, and its basis contains five Fe atoms, denoting by \textit{A},
\textit{B}, \textit{C}, \textit{D}, and \textit{E}. The \textit{E} sites in
the 245 layer are vacancy positions, denoted by empty circles.}
\label{fig1}
\end{figure}

We start by modeling the phase-separated region of K$_{y}$Fe$_{2-x}$Se$_{2}$
as a bilayer junction \cite{Jiang2012}, shown in Fig. \ref{fig1}. Here the
top layer is nominally taken as the SC 122 phase and the bottom layer is the
AF 245 phase. The system is governed by the Hamiltonian,
\begin{equation}
H=H_{122}+H_{245}+H^t_{\perp }+H^J_{\perp}.  \label{H_total}
\end{equation}%
Here $H_{122}$ and $H_{245}$ are the individual Hamiltonian for the 122 and
245 layers, $H^t_{\perp }$ is the Hamiltonian for interlayer hopping, and 
$H^J_{\perp }$ is the Hamiltonian for the interlayer 
spin coupling. To include
multi-orbital effect, we shall focus on the most relevant orbitals by
considering two orbitals only, $d_{\overline{x}\overline{z}}$ and $d_{%
\overline{y}\overline{z}}$ with $\overline{x}\ $or $\overline{y}$\ being
along the nearest neighbor Fe-Fe direction. In the following, $c_{\tau }$ ($%
d_{\tau }$) denotes the electron annihilation operator for the 122 (245)
layer with $\tau =1$ and $2$ representing $d_{\overline{x}\overline{z}}$ and
$d_{\overline{y}\overline{z}}$, respectively. All the energies are in unit
of electron-volt (eV).

For the 122 layer, $H_{122}$ contains a hopping term and a pairing term. The
hopping term is described by Das and Balatsky \cite{Das2011}, which yields
FS pockets at ($\pi ,0$) and ($0,\pi $) in the 1Fe/cell picture. The pairing
term with an attractive potential within nearest neighbor and next nearest
neighbor sites is given by
\begin{align}
H_{\Delta }& =-V_{1}\textstyle\sum\limits_{i,\bar{d}=\bar{x},\bar{y}}%
\textstyle\sum\limits_{\tau ,\sigma }c_{\tau ;i+\bar{d},\sigma }^{\dag
}c_{\tau ;i,-\sigma }^{\dag }c_{\tau ;i,-\sigma }c_{\tau ;i+\bar{d},\sigma }
\label{Hpair} \\
& -V_{2}\textstyle\sum\limits_{i,\bar{d}=\bar{x}\pm \bar{y}}\textstyle%
\sum\limits_{\tau ,\sigma }c_{\tau ;i+\bar{d},\sigma }^{\dag }c_{\tau
;i,-\sigma }^{\dag }c_{\tau ;i,-\sigma }c_{\tau ;i+\bar{d},\sigma }.  \notag
\end{align}%
Here $\sigma $ is the spin index and $V_{1}$ and $V_{2}$ are positive.

To describe the 245 layer, we note that the unit cell with one vacancy
contains 8 orbitals of electrons. The tight-binding model that respects the $%
I/4m$ symmetry of $\sqrt{5}\times \sqrt{5}$ Fe-vacancy order is constructed
in Ref. \onlinecite{Huang2011}. Following Ref. \onlinecite{Huang2011}, $%
H_{245}$ is constructed by removing $d_{\overline{x}\overline{y}}$ with
parameters being modified. The parameters for hoppings are $t_{11,\overline{x%
}}=-t_{11,\overline{y}}=0.3$, $t_{11,\overline{x}}^{\prime }=0.2$, $t_{11,%
\overline{y}}^{\prime }=0.15$, $t_{11,\overline{x}+\overline{y}}=-0.15$, $%
t_{11,\overline{x}-\overline{y}}=t_{11,\overline{x}+\overline{y}}^{\prime
}=-t_{11,\overline{x}-\overline{y}}^{\prime }=-0.08$, $t_{12,\overline{x}
}=t_{12,\overline{y}}=t_{12,\overline{x}}^{\prime }=0$, $t_{12,\overline{x}+
\overline{y}}=t_{12,\overline{x}+\overline{y}}^{\prime }=t_{12,\overline{x}-
\overline{y}}^{\prime }=-0.02$, and $\Delta =0.08$. Here $t_{\tau
\tau ^{\prime },\overline{R}}$ ($t_{\tau \tau ^{\prime },\overline{
R}}^{\prime }$) are for intracell (intercell) hoppings between
orbitals $\tau $ and $\tau ^{\prime }$ along $\overline{
R}$ directions and $\Delta$ is the site energy
difference between $d_{\overline{x}\overline{z}}$ and $d_{
\overline{y}\overline{z}}$ orbitals. Other hopping parameters can
be obtained from above by using symmetries of the system, for example, $t_{22,\overline{x}}=t_{11,
\overline{y}}$ as a result of the 4-fold rotational symmetry. In addition to hopping, the interaction between
electrons in the 245 layer is given by the generalized Hubbard model\cite{Huang2011}
\begin{align}
H_{I}& =\underset{i}{{\textstyle\sum }}\underset{I=A,B,C,D}{{\textstyle\sum }
}\left\{ U\underset{\tau }{{\textstyle\sum }}n_{\tau I,i\uparrow
}^{(d)}n_{\tau I,i\downarrow }^{(d)}\right.  \notag \\
& +\left[ \left( U^{\prime }-\frac{J_{H}}{2}\right)
n_{1I,i}^{(d)}n_{2I,i}^{(d)}-2J_{H}\mathbf{S}_{1I,i}^{(d)}\cdot \mathbf{S}%
_{2I,i}^{(d)}\right.  \notag \\
& \left. \left. +J_{C}\left( d_{1I,i\uparrow }^{\dag }d_{1I,i\downarrow
}^{\dag }d_{2I,i\downarrow }d_{2I,i\uparrow }+h.c.\right) \right] \right\} .
\label{H_I}
\end{align}%
Here \textit{A}, \textit{B}, \textit{C}, and \textit{D} denote Fe atoms in
the unit cell, as illustrated in Fig. \ref{fig1}. The onsite interaction
parameters follow the relations, $U^{\prime }=U-2J_{H}$, $J_{C}=J_{H}$, and $%
J_{H}=0.2U$. The chemical potential is used to control the particle density
per iron at $n=2$.

As two layers couple, the unit cell of the 122 layer is enlarged as that of
the 245. We shall denote Fe atoms by \textit{A, B, C, D}. Here \textit{E} is
the position of the vacancy in 245, as illustrated in Fig. \ref{fig1}. $
H^t_{\perp }$ is given by
\begin{equation}
H^t_{\perp }=t_{\perp }\textstyle\sum\limits_{i}\textstyle\sum%
\limits_{I=A,B,C,D}\textstyle\sum\limits_{\tau ,\sigma }\left( c_{\tau
I;i\sigma }^{\dag }d_{\tau I;i\sigma }+d_{\tau I;i\sigma }^{\dag }c_{\tau
I;i\sigma }\right) .
\end{equation}
On the other hand, the interlayer spin interaction $H^J_{\perp }$ is given by
\begin{equation}
H^J_{\perp }=J_{\bot }\textstyle\sum\limits_{i}\textstyle
\sum\limits_{I=A,B,C,D} \textstyle\sum\limits_{\tau ,\tau ^{\prime }}
\mathbf{S}_{\tau I,i}^{(c)}\cdot \mathbf{S}_{\tau ^{\prime }I,i}^{(d)}.
\end{equation}
Here $\mathbf{S}_{\tau I,i}^{(c)}$ is the spin operator of electrons in the 122 layer, while
$\mathbf{S}_{\tau ^{\prime }I,i}^{(d)}$ is that of electrons in the 245 layer. For Fe-based superconductors,
$J_{\bot }$ is found to be in the range of $1-5$ meV\cite{Jperp}. We shall set $J_{\bot}$ to be a nominal value
of $10$ meV in this work.
\begin{figure}[tbp]
\begin{center}
\includegraphics[height=2.1in,width=3.5in] {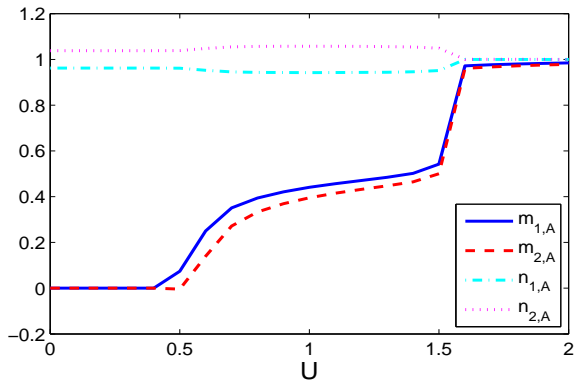}
\end{center}
\caption{(Color online) Mean field solutions of the checkerboard AF states
of $H_{2}$ for the Hubbard \textit{U} being up to two. Magnetic moments ($%
m=n_{\uparrow }-n_{\downarrow }$) and particle densities ($n=n_{\uparrow
}+n_{\downarrow }$) at site-\textit{A} and \textit{B} for orbital-1 and 2
are respectively displayed. The antiferromagnetism arises at the critical
point $U=0.5$, and it jumps to being fully magnetized at $U=1.6$.}
\label{fig2}
\end{figure}

\begin{figure}[tbp]
\begin{center}
\includegraphics[height=2.6507in,width=3.4229in] {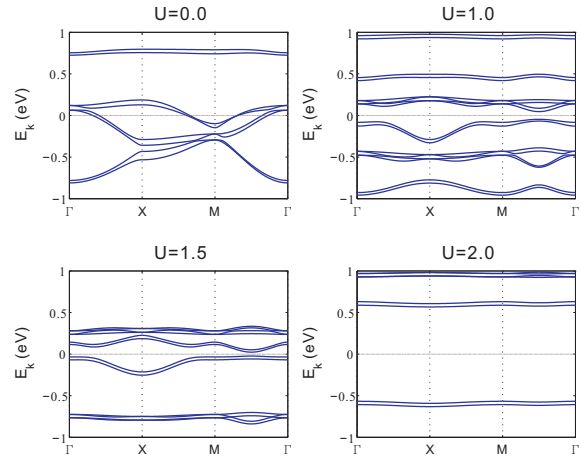}
\end{center}
\caption{Energy dispersions in the range $(-1,1)$ of the AF states for $%
H_{245}$ at \textit{U}=0, 1.0, 1.5, and 2.0, respectively. It is clear that
above \textit{U} $\approx $1.5, a large AF gap, $\Delta _{AF}$, opens. }
\label{fig3}
\end{figure}

In the following, we shall first turn off $J_{\perp}$ and consider
effects of $t_{\perp}$. The distance between two nearest vacancies will be set as
unity and their directions are denoted as \textit{x} and \textit{y}. For the isolated
245 layer, the mean field solutions of the AF order are shown in Fig. \ref
{fig2}. Since we shall focus on effects of AF order on superconductivity, AF orders are treated as 
boundary conditions and will not be solved self-consistently later when interlayer
couplings are turned on. Hence these values obtained in Fig. \ref{fig2} will be adopted later even when the interlayer couplings
are turned on. It is seen that the antiferromagnetism is
weak when $U\lessapprox 1.5$ and is strong with saturated magnetization when
$U>1.5$. In Fig. \ref{fig3}, we show the energy dispersions of the AF states
with \textit{U}=0, 1.0, 1.5, and 2.0, at which the development of the
antiferromagnetism is at the beginning, in the middle, right before the
jump, and at the saturation, respectively. Clearly, $H_{245}$ reproduces the
quasi-nested Fermi surface (FS) and the expected block checkerboard
antiferromagnetism [with $\mathbf{q}=(\pi ,\pi )\equiv \mathbf{Q}$] as the
Hubbard \textit{U} increases above the critical value, $U\approx 0.5$. The
minimal gap is along the diagonal ($\Gamma -M$ ) direction and is smaller
than 0.05eV before the saturation. When $U>1.5$, a large AF gap, $2\Delta
_{AF}$, opens. However, for $1.0<U<1.5$, we find that although bands are
expelled to higher energy as \textit{U} increases, the energy gap near the
chemical potential decreases, resulting in the AF gap at \textit{U}
=1.0 ($\Delta _{AF}$=0.046) is larger than that at \textit{U}=1.5 ($%
\Delta _{AF}$=0.023). Note that with two \textit{d}-orbitals, one can not
produce detailed characteristics of the band structure as those obtained by
the first principal calculations \cite{Yan2011,Huang2011}. However,
important relevant features are reproduced with the AF state being an gapped
insulator and the AF order being agrees with the experimental observation.

\begin{figure*}[tbp]
\centering{\includegraphics[width=6.89in] {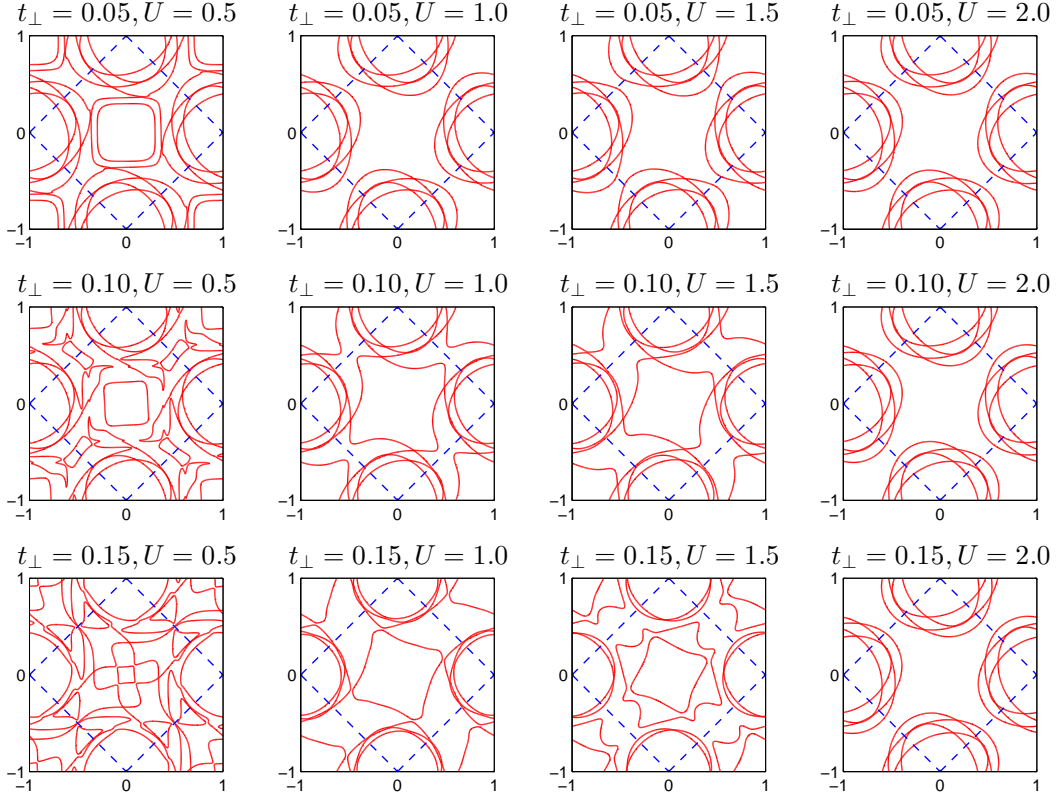} }
\caption{(Color online) The FS contours in the absence of superconductivity
when 122-245 two layers couple together. Panels in the first row are for $%
t_{\perp }$=0.05; those in the middle are for $t_{\perp }$=0.1; those in the
last are for $t_{\perp }$=0.15. While panels in the first column are for
\textit{U}=0.5, second for \textit{U}=1.0, third for \textit{U}=1.5, and
fourth for \textit{U}=2.0. Unit of the \textit{x}-axis is $k_{x}/\protect\pi
$ and that of the \textit{y}-axis is $k_{y}/\protect\pi $. Here dashed lines
are the magnetic Brillouin zone boundaries and the AF order parameters are
those obtained from the isolated 245 layer. }
\label{fig4}
\end{figure*}

After the interlayer hopping is included, FS's of the 122 layer start to
deform as shown in Fig. \ref{fig4}. Here three values of $t_{\perp }$ ($%
t_{\perp }$=0.05, 0.1, and 0.15) for \textit{U}=0.5, 1.0, 1.5, and 2.0 are
shown. Panels in the first row are for $t_{\perp }$=0.05, the second row are
for $t_{\perp }$=0.1, and the third one are for $t_{\perp }$=0.15 On the
other hand, columns from the left to the right are cases with \textit{U}%
=0.5, \textit{U}=1.0, \textit{U}=1.5, and \textit{U}=2.0, respectively. It
is seen that for large $U$ with large AF orders, FS sheets at ($\pi $,0) are
always disconnected from those at (0,$\pi $); while for weak AF orders,
interlayer hopping deforms FS pockets at ($\pi $,0) and (0,$\pi $) so that
they start to connect with each other and, consequently, more FS pockets
emerge and the electronic structure become very complicated. In particular,
as $t_{\perp }$ increases or \textit{U} decreases, apparent FS pockets
emerge around $\Gamma $ and M, which are mainly contributions from the AF
state.

\section{Results}

We first characterize SC states in the 122 system. For this purpose, we note
that since $A_{g}$ (\textit{s}-wave) and $B_{g}$ (\textit{d}-wave) are the
two major competing order parameters \cite{Huang2012}, we shall only
consider these pairing symmetries. In addition, in constructing SC order
parameters, one needs to impose point group symmetries. Therefore, the $d_{%
\overline{y}\overline{z}}$ -orbital pairs are obtained from the $d_{%
\overline{x}\overline{z}}$-orbital pairs, \textit{e.g.}, $\left\langle
c_{2A;i}c_{2B;i}\right\rangle =\pm \left\langle
c_{1B;i}c_{1C;i}\right\rangle $, where $\pm $ denotes \textit{s}-wave/%
\textit{d}-wave. In addition, due to the presence of vacancies, translation
symmetry will not hold always, \textit{e.g.}, $|\left\langle
c_{1D;i}c_{1E;i}\right\rangle |\neq |\left\langle
c_{1A;i}c_{1D;i}\right\rangle |$.
\begin{figure}[tbp]
\includegraphics[width=3.45in]{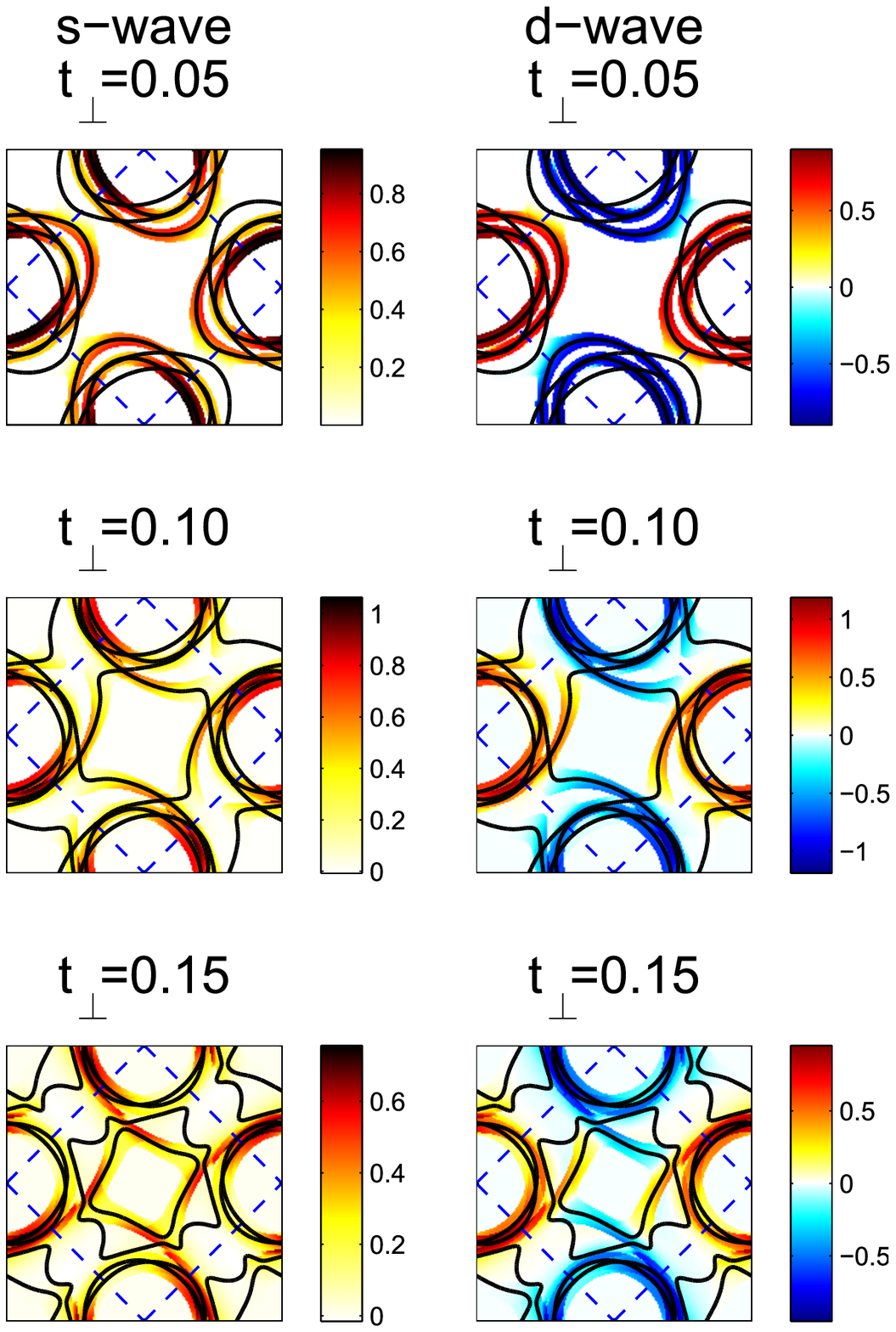}\newline
\caption{(Color online) Gap functions $\Delta (\mathbf{k})$ of \textit{s}%
-wave (left column) and of \textit{d}-wave (right column) near FS's at $%
T_{c} $. Three cases of $t_{\perp }$'s are compared: $t_{\perp }=0.05$, $%
t_{\perp }=0.10$, and $t_{\perp }=0.15$. Other common parameters are $%
V_{1}=0.175$, $V_{2}=0.225$, and $U=1.5$. The scale is arbitrary.}
\label{fig5}
\end{figure}

Given the pairing symmetry, $T_{c}$ is obtained by solving the
Bethe-Salpeter equation
\begin{equation}
\Delta ^{ij}(K)=-\lambda \textstyle{\sum\limits_{K^{\prime }}}\Delta
^{m^{\prime }n^{\prime }}(K)G_{mm^{\prime }}(K^{\prime })G_{nn^{\prime
}}(-K^{\prime })V_{mn}^{ij}(K,K^{\prime }).  \label{BS-eq1}
\end{equation}%
Here $K=(\mathbf{k},i\omega _{n})$ with $\omega _{n}=(2n+1)\pi kT$ being the
Matsubara frequency, the indices $i$ and $j$ are orbital-sublattice indices containing 
both orbital and site labels, and implicit summation over orbital indices $m,n,m^{\prime }$, and $n^{\prime }$
is taken. $\Delta $ is the pairing amplitude, $\lambda $ is the
eigenvalue, and \textit{V} is the pairing interaction. $G_{mm^{\prime }}$
is the Green's function with orbital indices $m,m^{\prime }$ defined by
\begin{equation}
G_{mm\prime }(\mathbf{k},i\omega _{n})=\textstyle\sum\limits_{\mu }\frac{%
A_{m\mu }(\mathbf{k})A_{m^{\prime }\mu }^{\ast }(\mathbf{k})}{i\omega
_{n}-\xi _{\mu }(\mathbf{k})}.
\end{equation}%
Here $\xi _{\mu }$ is the energy of the band $\mu $. $A_{m\mu }$ is the
transformation matrix that connects the orbital basis $\psi _{m}(\mathbf{k})$
to the eigen-energy basis $\gamma _{\mu }(\mathbf{k})$ via the relation $
\psi _{m}(\mathbf{k})=\sum_{\mu }A_{m\mu }(\mathbf{k})\gamma _{\mu }(\mathbf{%
k})$. It is more convenient to work in the $k$-space by transforming Eq. (%
\ref{BS-eq1}) into band representation using the transformation matrix $%
A_{m\mu }$. We shall assume that pairing is among intra-band and decompose
the interaction into different bases $g_{a}(\mathbf{k})$
\begin{equation}
V_{mn}^{ij}(\mathbf{k},\mathbf{k}^{\prime })=-\delta _{im}\delta _{jn}%
\textstyle\sum\limits_{a}\mathcal{V}_{a}^{ij}g_{a}(\mathbf{k})g_{a}^{\ast }(%
\mathbf{k}^{\prime }).
\end{equation}%
By multiplying both sides of Eq. (\ref{BS-eq1}) by $A_{i\mu }(-\mathbf{k}%
)A_{j\mu }(\mathbf{k})$ and summing over \textit{i}, \textit{j} and then
performing the Matsubara frequency summation, Eq. (\ref{BS-eq1}) is
transformed into the representation in the band basis
\begin{equation}
\Delta _{\mu }(\mathbf{k})=2\textstyle\sum\limits_{a^{\prime }}\textstyle%
\sum\limits_{i^{\prime }\leq j^{\prime }}\Re \left[ g_{a^{\prime }}(\mathbf{k%
})A_{i^{\prime }\mu }(-\mathbf{k})A_{j^{\prime }\mu }(\mathbf{k})\right]
\mathcal{J}_{a^{\prime }}^{i^{\prime }j^{\prime }},  \label{delta}
\end{equation}%
with $\mathcal{J}_{a}^{ij}$ being the order parameter satisfying a
self-consistent equation
\begin{equation}
\mathcal{J}_{a}^{ij}=\lambda \textstyle\sum\limits_{a^{\prime }}\textstyle%
\sum\limits_{i^{\prime }\leq j^{\prime }}\mathcal{V}_{a}^{ij}\mathcal{K}%
_{a,a^{\prime }}^{ij,i^{\prime }j^{\prime }}\mathcal{J}_{a^{\prime
}}^{i^{\prime }j^{\prime }}.  \label{BS-eq3}
\end{equation}%
Here $\Delta _{\mu }(\mathbf{k})\equiv \sum_{m,n}A_{m\mu }(-\mathbf{k}%
)A_{n\mu }(\mathbf{k})\Delta ^{mn}(\mathbf{k})$ and $\mathcal{K}_{aa^{\prime
}}^{ij,i^{\prime }j^{\prime }}$ is given by
\begin{eqnarray}
\mathcal{K}_{aa^{\prime }}^{ij,i^{\prime }j^{\prime }} &=&\frac{2}{N}%
\textstyle\sum\limits_{\mu }\textstyle\sum\limits_{\mathbf{k}}\Re \left[
g_{a}(\mathbf{k})A_{i\mu }(-\mathbf{k})A_{j\mu }(\mathbf{k})\right]
\label{susceptibility} \\
&&\times \Re \left[ g_{a^{\prime }}(\mathbf{k})A_{i^{\prime }\mu }(-\mathbf{k%
})A_{j^{\prime }\mu }(\mathbf{k})\right] \chi _{\mu }(\mathbf{k})  \notag
\end{eqnarray}%
with $\chi _{\mu }(\mathbf{k})\equiv \tanh \left( \xi _{\mu }(\mathbf{k}%
)/2kT\right) /2\xi _{\mu }(\mathbf{k})$. The SC state is found by solving
eigenvalues and eigenvectors of the vertex $\mathcal{\hat{V}\hat{K}}$ in Eq.
(\ref{BS-eq3}) with $T_{c}$ being obtained when $\lambda =1$.
\begin{figure}[tbp]
\includegraphics[width=3.5in]{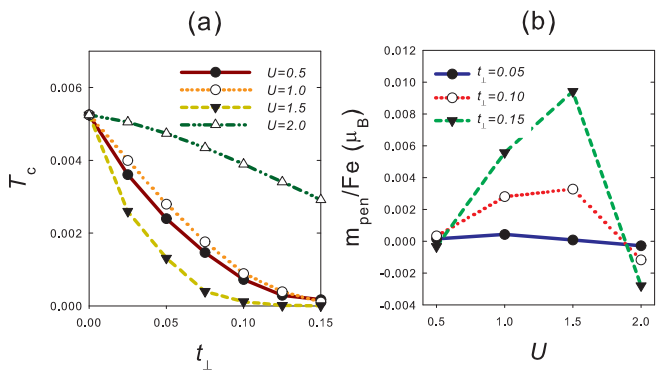}\newline
\caption{(Color online) (a) Suppression of $T_{c}$ versus $t_{\perp }$ of
\textit{s}-wave for different $U$'s. Here $V_{1}=0.175$ and $V_{2}=0.25$.
Note that $T_{c}$ of \textit{d}-wave has similar trend but the suppression
is more severe. (b) Average moment that penetrates into the 122 layer in the
normal state for different $U$ and $t_{\perp }$.}
\label{fig6}
\end{figure}

In Fig. \ref{fig5}, we examine the SC gap functions near the FS's defined by
\begin{equation}
\Delta (\mathbf{k})=\textstyle\sum\limits_{\mu }\Delta _{\mu }(\mathbf{k}%
)\Theta (\epsilon -\left\vert \xi _{\mu }(\mathbf{k})\right\vert ).
\end{equation}%
Here $\epsilon $ is a small energy cutoff that restricts the gap function to
be exhibited near FS. Three different $t_{\perp }$, 0.05, 0.10, and 0.15 at
\textit{U}=1.5 are shown for \textit{\ s}-wave in the left column and
\textit{d}-wave in the right column. Clearly, one sees that as the
interlayer hopping increases and the FS contours start to deform, the gap
functions becomes more anisotropic and the condensation energy decreases.
However, the characters of both \textit{s}-wave and \textit{d}-wave are
clearly kept on the deformed FS as shown in Fig. \ref{fig5} for $t_{\perp
}=0.15$, where zeros of $\Delta (\mathbf{k})$ are present for the \textit{d}%
-wave on the central FS, but they do not exist for the \textit{s}-wave. This
symmetry property makes the \textit{d}-wave more disadvantageous than the
\textit{s}-wave when two layers are strongly coupled. Hence \textit{d}-wave
is more sensitive to the interlay coupling than the \textit{s}-wave. Note
that since there is only an indirect correlation between particles with
momentum $\mathbf{k}$ and $\mathbf{k+Q}$ via the 245 layer, $\Delta (\mathbf{%
k})$ does not have to be equal to $\Delta (\mathbf{k+Q})$.

We now examine effects of the interlayer hopping and the AF order on the SC
transition. First, we note that similar to the situations in iron-pnictides
\cite{Seo2008}, $V_{1}$ tends to favor the \textit{d}-wave while $V_{2}$
favors the \textit{s}-wave, and the phase boundary is at $V_{1}/V_{2}\sim
0.78$. To be concrete, we shall set $V_{1}=0.175$ and $V_{2}=0.25$ and focus
on $T_{c}$ of the \textit{s}-wave. Similar behavior is found for \textit{d}%
-wave. Fig. \ref{fig6}(a) shows $T_{c}$ of \textit{s}-wave SC order versus $%
t_{\perp }$ for different values of $U$. Clearly, for a given AF order
(fixed by $U$), it is seen that $T_{c}$ always gets suppressed by $t_{\perp
} $. However, for fixed $t_{\perp }$, when $U$ increases, the change of $%
T_{c}$ is non-monotonic (due to non-monotonic $\Delta _{AF}$) 
and $T_{c}$ is only weakly suppressed at $U\sim 2$. Further analysis
shown in Fig. \ref{fig6}(b) indicates that the penetrated AF order into the
122 layer has the inverse trend as that of $T_{c}$. These behaviors can be
understood by examining FS structures. Comparison of Fig. \ref{fig4} and
Fig. \ref{fig6}(a) shows that the suppression of the SC order is due to the
deformed FS structures induced by the interlayer hopping. The deformed FS
structures generally frustrate the coupling of SC orders on FS's and thus
suppress the SC order. However, in the presence of strong AF order, the 245
layer is insulating with a gap. Since the coherence length $\xi _{AF}$ of an
AF phase is $\xi _{AF}\sim \frac{\hbar v_{F}}{\Delta _{AF}}$ with $v_{F}$
being the characteristic Fermi velocity, a large AF gap implies a short
penetration depth of the AF order into the SC layer. Hence the induced
deformation of FS structure is weak, which leads to weak suppression of
superconductivity. Note that since the interlayer hopping between AF layer
and SC layer suppresses $T_{c}$, these results imply that the pure
SC phase has a higher $T_{c}$. If one takes $t_{\perp }=0.15$ and $U=2$ as a
reasonable estimation of phase-separated ternary iron selenides, the real SC
transition is around $65K$, which is comparable to highest observed $T_{c}$
in the family of iron selenides \cite{Xue2012}. The suppression of $
T_{c}$ due to the interlayer hopping $t_{\perp }$ is also
studied by Berg \textit{et al}. \cite{Berg2009} for a one-band negative \textit{U} model, 
in which it is shown that the leading order
correction to the pairing susceptibility is negative and is proportional to $t_{\perp }^{2}$. 
As a result, in their model, $T_{c}$ is suppressed by the order of $
t_{\perp }^{2}$ for small $t_{\perp }$. The susceptibility suppression
also happens in our case as one can see that in Eq. (\ref{susceptibility})$, 
t_{\perp }$ will change $\xi _{\mu }$ and $A_{i\mu }$ and thus values of $
\mathcal{K}$. However, due to multi-orbital nature of our model, 
the behavior of $T_{c}$\ at small $t_{\perp }$ do not follow simple quadratic behavior.
Only for weak suppression of $T_{c}$ at large \textit{U} shown in \ref{fig6}(a), 
we find that suppression of $T_{c}$ is quadratic in $t_{\perp}$, in agreement with
results found in Ref. [\onlinecite{Berg2009}].

\begin{figure}[tbp]
\includegraphics[width=3.4in]{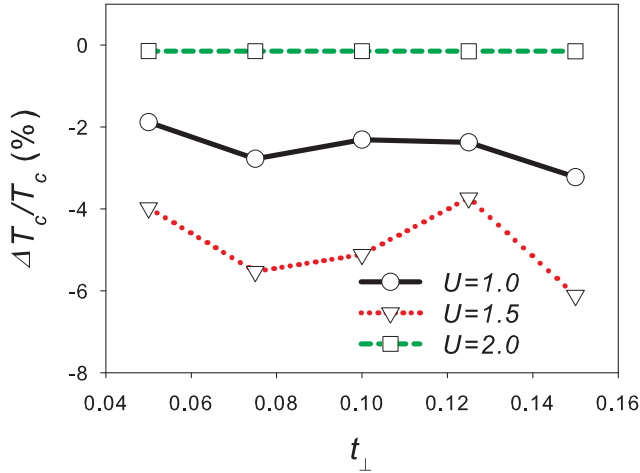}\newline
\caption{(Color online) Suppression of \textit{s}-wave $T_{c}$ ($\Delta T_{c}/T_{c}$) versus $t_{\perp}$ 
for a given interlayer spin interaction $J_{\perp}$ under three different \textit{U}'s in the
245 system. Here $V_{1}=0.175$ and $V_{2}=0.25$ are same as those
adopted in Fig. \protect\ref{fig6}. The interlayer spin coupling $J_{\perp }$ is 0.01 with
the corresponding $V_{J}$'s for \textit{U}=1.0, 1.5, and 2.0 cases being
0.0016, 0.0032, and 0.0001 respectively.}
\label{fig7}
\end{figure}

Finally we examine effects of interlayer spin interaction $H^J_{\perp }$ on
superconductivity.  For this purpose, we first note that $H^J_{\perp }$ only
characterizes single particle scattering in the 122 layer and 245 layer respectively.
Hence the effective Hamiltonian for scatterings of Cooper pairs in the 122 layer 
must be second order in $H^J_{\perp }$. To the second order in the
perturbation theory, scattering of two particle for particle-particle
channel in the 122 layer is given by
$T_{\perp
}^{(2)} \equiv H^J_{\perp } \left( E_0-H_{0}\right) ^{-1} H^J_{\perp }$, where $E_0$ is the unperturbed ground state
energy, $H_{0}=H_{122}+H_{245}$, and $H_0-E_0$ is the energy excitation for the intermediate
state. During the scattering of two particles in the 122 layer, scatterings 
in the 245 layer are captured by the magnetic susceptibility with the major weight being in particle-hole excitations
Since the particle-hole excitation energy of an AF insulator is the sum of the energies 
for two quasi-particles above the AF gap, the change of energy
for the intermediate state during scattering of Cooper pairs is at least  $2\Delta _{AF}$.
By neglecting dispersion of energy spectrum, we
find $\left( E_0-H_{0}\right) ^{-1}\approx $ $-(2\Delta _{AF})^{-1}$.
Therefore, after taking average over \textit{d} electrons of the 245 layer,  
the effective intra-orbital pairing Hamiltonian due to interlayer spin interaction
is given by
\begin{equation}
\delta H_{\Delta }=V_{J}\textstyle\sum\limits_{i,\bar{d}=\bar{x},\bar{y},%
\bar{x}\pm \bar{y}}^{\prime }\textstyle\sum\limits_{\tau ,\sigma }c_{\tau ;i+%
\bar{d},\sigma }^{\dag }c_{\tau ;i,-\sigma }^{\dag }c_{\tau ;i,-\sigma
}c_{\tau ;i+\bar{d},\sigma },
\end{equation}%
where the summation does not include \textit{E} sites and $V_{J}=%
\frac{3J_{\bot }^{2}}{4\Delta _{AF}}$. We note that it is a repulsive interaction for
Cooper pairs and hence the interlayer interaction tends to suppress
superconductivity. In Fig. \ref{fig7}, we examine changes of $T_{c}$ for $J_{\perp }$
=0.01 in three different \textit{U}'s with corresponding $V_{J}$ being
0.0016, 0.0032, and 0.0001.  It is seen that similar to effects of $t_{\perp}$, 
$T_c$ gets suppressed but the suppression is non-monotonic and the
variation of suppression is less than the suppression due to different \textit{U}'s.
In particular, similar to the suppression by $t_{\perp}$, a stronger AF phase gets less 
suppression in superconductivity. The mechanism behind the behavior
of suppression of $T_c$ is clearly due to the dependence of effective pairing
strength $V_{J}$ being inversely proportional to $\Delta _{AF}$. In addition to 
direct interlayer spin coupling, in real materials, $J_{\perp }$ may arise from super-exchange interaction
between the 122 and 245 layers. In that situation, 
$J_{\perp }$ is proportional
to $\frac{t_{\bot }^{2}}{U}$. Since $U \sim \Delta_{AF}$, we have $J_{\perp}
\sim \frac{t_{\bot }^{2}}{\Delta _{AF}}$. As a result, not only the wave-function hybridization 
due to $t_{\bot }$ but also the interlayer spin interaction resulted from $t_{\bot}$ suppress 
pairing and result in non-monotonic suppression of $T_{c}$ with $T_c$ being less suppressed 
for large $\Delta _{AF}$. Hence while both interlayer hopping and interlayer spin couplings suppress superconductivity,
a large AF order in the 245 layer can result in stronger superconductivity in the 122 layer.

\section{Summary}

In summary, we have found that the existence of large magnetic moment in the
AF phase is the key reason of why the phase-separated ternary iron selenides
can maintain a relative high $T_{c}$ despite of the strong competition
between SC and AF orders. Based on a minimal bilayer model with both the 122
and 245 phases, we show that proximity effects of the AF order on the SC
order generally result in the deformation of FS due to the interlayer
hopping and Cooper pair scatterings due to the interlayer spin interaction. 
It is shown that the deformed FS's generally frustrate coupling of SC orders and result in
the suppression of superconductivity. In addition,  the interlayer spin coupling generates 
repulsive Cooper pair scattering and it also tends to suppress superconductivity.
However, when the AF phase has a large AF
order, it is insulating with a large gap and the penetration of the AF order
into the SC layer is suppressed. As a result, the superconductivity is
protected against interlayer hopping and interlayer spin coupling 
when $\Delta _{AF}$ is much larger than the interlayer hopping.

While our results are consistent with experimental observations made so far,
there are a number of experimental observations of 3D-like FS's in the
phase-separated region \cite{Yan2011,Cao2011,ZHLiu2012}. To account for
these experimental results, it would require a relatively large interlayer
coupling. Since the interlayer hopping between AF and SC layers suppresses $%
T_c$, our results imply that 2D-like system may be more preferable for
higher $T_c$. In fact, the real SC phase in phase-separated ternary iron
selenides may have a higher $T_c$ up to $65K$ , which is in comparable to
highest observed $T_c$ in the family of iron selenides \cite{Xue2012b}.

\begin{acknowledgments}
We thank Prof. Ming-Che Chang for useful discussions. This work was
supported by the National Science Council of Taiwan.
\end{acknowledgments}

\end{document}